\documentclass[twocolumn]{aa}

\usepackage{txfonts}
\usepackage{graphicx}
\usepackage{natbib}
\usepackage{mathrsfs}
\bibpunct{(}{)}{;}{a}{}{,}

\begin{document}

\title{Shells, jets, and internal working surfaces in the molecular outflow 
from IRAS 04166+2706\thanks{Based on observations carried out with the 
IRAM Plateau de Bure Interferometer. 
IRAM is supported by INSU/CNRS (France), MPG (Germany) and IGN (Spain).}}

\author{J. Santiago-Garc\'{\i}a \inst{1}
\and
M. Tafalla \inst{1}
\and D. Johnstone \inst{2}
\and R. Bachiller \inst{1}
}

\institute{Observatorio Astron\'omico Nacional (IGN),
Alfonso XII 3, E-28014 Madrid,
Spain
\and
NRC Canada, Herzberg Institute of Astrophysics,
5071 West Saanich Road, Victoria, B.C. V9E 2E7, Canada
}

\offprints{M. Tafalla \email{m.tafalla@oan.es}}
\date{Received -- / Accepted -- }

\abstract
{IRAS 04166+2706 in Taurus is one of the most nearby young stellar objects
whose molecular outflow contains a highly collimated fast component. 
}
{The high symmetry and pristine appearance of this outflow suggests that its 
study could offer unique clues on the nature of the still mysterious 
underlying driving wind.}
{We have observed the IRAS 04166+2706 outflow with the IRAM Plateau de Bure 
interferometer in CO(J=2--1) and
SiO(J=2--1) achieving angular resolutions between $2''$ and $4''$. To improve the
quality of the CO(2--1) images, we have added single dish data to the 
interferometer visibilities.}
{The outflow consists of two distinct components. At
velocities $<10$~km~s$^{-1}$, the gas forms two opposed, approximately 
conical shells that have the YSO at their vertex. These shells coincide
with the walls of evacuated cavities and seem to result from the acceleration
of the ambient gas by a wide-angle wind. At velocities $>30$~km~s$^{-1}$, 
the gas forms two opposed jets that travel along the center
of the cavities and whose emission is dominated by a 
symmetric collection of at least 7 pairs of peaks. The velocity field 
of this component presents a sawtooth pattern with the
gas in the tail of each peak moving faster than the gas in the head. 
This pattern, together with a systematic widening of the peaks
with distance to the central source, is consistent with the emission 
arising from internal working surfaces traveling along the
jet and resulting from variations in the velocity field of ejection. We interpret
this component as the true protostellar wind, and we find its composition
consistent with a chemical model of such type of wind.}
{Our results support outflow wind models that have simultaneously wide-angle and
narrow components, and suggest that the EHV peaks seen in a number of
outflows consist of internally-shocked wind material.}


\maketitle

%

\section{Introduction}

Bipolar outflows powered by young
stellar objects (YSOs) still pose a number of puzzles
almost three decades after their discovery \citep{sne80}. 
Although it is now well accepted that most of the moving gas seen 
in an outflow consists of molecular 
ambient material accelerated by a  
primary wind, inferring the properties of this wind from 
observations of the accelerated gas has proven a difficult task 
\citep[e.g.,][]{bac96}. 
Part of this difficulty results from the large
variety of observed shapes and velocity patterns seen in the
outflow accelerated gas,
that range from the almost parabolic shells of low velocity material
seen in \object{L1551} and  \object{Mon R2} \citep{mor87,mey91}
to the collimated jets of fast gas
observed towards \object{L1448-mm} and \object{HH211} \citep{bac90,gue99}. 
In addition to the molecular data,
optical observations of outflows often reveal an atomic
component of higher excitation and collimation, probably resulting 
from recently shocked material either in the wind itself or at the interface
between the wind and the cloud. This atomic component
is clearly related to the molecular gas, as the two coincide 
in sense and direction when both are seen, but the exact connection 
between these components still remains unclear \citep{pet06}.
The presence of this component reveals that the outflow phenomenon 
is both long lived and highly time variable 
(see \citealt{rei01} for a review).

A number of wind geometries and kinematics
have been proposed over the years
to explain outflow observations, and most of the
models can be classified as either 
wide-angle components \citep[e.g.,][]{shu91} or
highly collimated jets \citep[e.g.,][]{mas93}. Each of these simple geometries 
has proven successful explaining a subset of the observed outflows:
wide-angle winds can easily explain shell outflows, while jets
have been used to model highly collimated flows.
None of these models, however, seems capable of explaining simultaneously
all types of outflows \citep{lee02}, and this suggests that each of them only
provides a partial description of the outflow phenomenon.

In recent years, a new generation of outflow models with  
both collimated and wide-angle wind components
has been presented \citep{ban06,sha06,mac08}. 
At the same time, a significant
increase in the sensitivity and resolution of millimeter
and submillimeter
interferometers such as the PdBI, SMA, or CARMA 
has started to allow mapping the relatively weak emission from outflows
with a detail approaching that of the optical and IR
observations of Herbig-Haro objects and jets. 
This combination of theoretical and observational advances has
opened the possibility of carrying out a new generation of 
outflow studies that may finally reach a consistent
picture of outflows. Such a picture should not only describe the 
variety of observed morphologies and kinematics, but 
bring together, in a single evolutionary sequence, 
both the molecular outflows
from the most embedded objects and the optical jets of
some TTauri stars. 

Previous studies of low-mass outflows have shown that those from 
the youngest stellar objects display the simplest geometry
and kinematics, probably due to their limited
distortion by anisotropies in the surrounding
environment \citep{arc07}. 
These youngest outflows often present in the spectra a distinct
secondary component at velocities over 30~km~s$^{-1}$, which is
often referred to as the extremely high velocity (EHV)
gas \citep{bac96}.
Because of their pristine appearance,
these young outflows are
ideal targets to attempt inferring properties of the
invisible wind from the observation of the accelerated
gas. Among the youngest protostars known, \object{IRAS 04166+2706}
(I04166 hereafter) stands out for its close
location (Taurus molecular cloud, at 140~pc), 
simple environment (the \object{B213} filament), and
high symmetry of its bipolar outflow. The flow from
this 0.4~L$_\odot$ object was first reported by \citet{bon96}, 
who mapped its low velocity gas towards the vicinity of the YSO.
Further observations with the IRAM 30m telescope by \citet{taf04}
(TSJB04 hereafter) revealed a spectacular outflow extending over 
at least $400''$ ($> 0.25$~pc) and presenting
a component of EHV gas that contains half of the outflow total
momentum and 80\% of its kinetic energy, and 
so collimated that could not be resolved
with a beam of $\approx 11''$ FWHM. 
In this paper, we present new high resolution interferometric 
observations of the central $\approx 0.1$~pc of the outflow.
These new data allow us to separate the distributions of the EHV gas
and the lower velocity component, and their analysis  
provides new clues on the nature of the outflow driving wind.

\section{Observations}

We observed the inner part ($\sim 200''\times 40''$) 
of the I04166 bipolar outflow 
using the IRAM Plateau de Bure Interferometer (PdBI)
in its CD configuration during the winters of 2004-05 (blue lobe)
and 2005-06 (red lobe). The receivers were tuned simultaneously
to SiO(J=2--1) (86.8 GHz) and CO(J=2--1) (230.5 GHz), 
and the correlator was configured to provide velocity
resolutions of 1.6 and 1.1~km s$^{-1}$ for the two transitions,
respectively. A broad band mode of the correlator was also used 
to provide 1~GHz of bandwidth at each frequency for making
maps of the continuum emission. In order to cover the
extended emission of the outflow, each lobe was observed with a 
mosaic of 18 fields, and the central two fields around I04166 
were observed in the two epochs to check and ensure 
the consistency of the calibration
(found to be better than 10\% for the continuum 
at both 1 and 3~mm).

The observations of I04166 were interspersed with observations of 
nearby quasars
to track possible variations of the instrumental gains 
during the observation. These calibrator observations were
used to correct the I04166 data, and the calibrated visibilities
were inverted and CLEANed to generate maps using the GILDAS 
software\footnote{http://www.iram.fr/IRAMFR/GILDAS}.
Comparing the resulting interferometer spectra with single dish
observations from the IRAM 30m telescope, we estimate that
the interferometer observations recover close to 100\% of the
flux from the SiO(2--1) emission. For CO(2--1), the 
interferometer data recover about 40\% of the flux at the highest
velocities and close to 20\% at the lowest speeds. To correct for
this flux loss, we used the CO(2--1) data presented in
TSJB04. From these data, we generated a set of short-spacing 
visibilities that were added to the PdBI visibilities, and a 
new series of maps was made. Before merging the data, 
we found that it was necessary to 
offset the single dish map by ($4''$, -$3''$)  
in order to align the two data sets.
Natural or robust weighting provided the best compromise
between resolution and sensitivity, and they were used 
accordingly to generate the final maps. 
The FWHM of the synthesized  beam is about $4'' \times 3''$ for SiO(2--1) 
and the 3.5~mm continuum and  $2'' \times 2''$ for CO(2--1) and 
the 1.3~mm continuum.

\section{Continuum data}

\begin{figure}
\centering
\resizebox{\hsize}{!}{\includegraphics[angle=270]{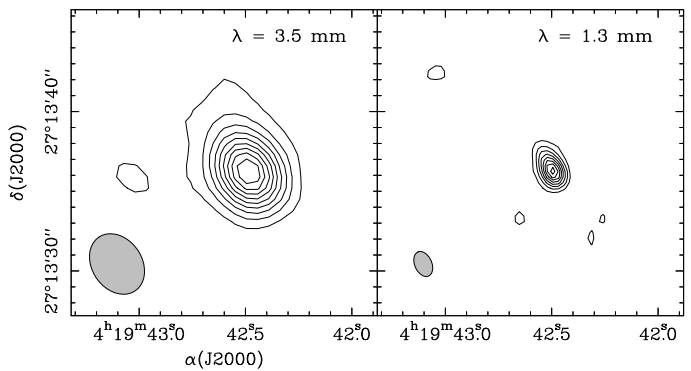}}
\caption{Continuum maps of the vicinity of IRAS 04166+2706
for wavelengths 3.5~mm (left) and 1.3~mm (right). Only the central part of
the map is shown because no additional mm sources were detected in the
region (see Fig.~\ref{fig_co_3vel} for the full extent of the map).
First contour and contour interval
is 1~mJy~beam$^{-1}$ for 3.5~mm and 7.5~mJy~beam$^{-1}$ for 1.3~mm.
The FWHM of the synthesized beam, indicated with a grey ellipse in each panel,
is $4\farcs1\times3\farcs1$ at 3.5~mm and $1\farcs7 \times 1\farcs1$ at 1.3~mm.
\label{fig_cont}}
\end{figure}

\begin{figure}[h!]
\centering
\resizebox{\hsize}{!}{\includegraphics[angle=270]{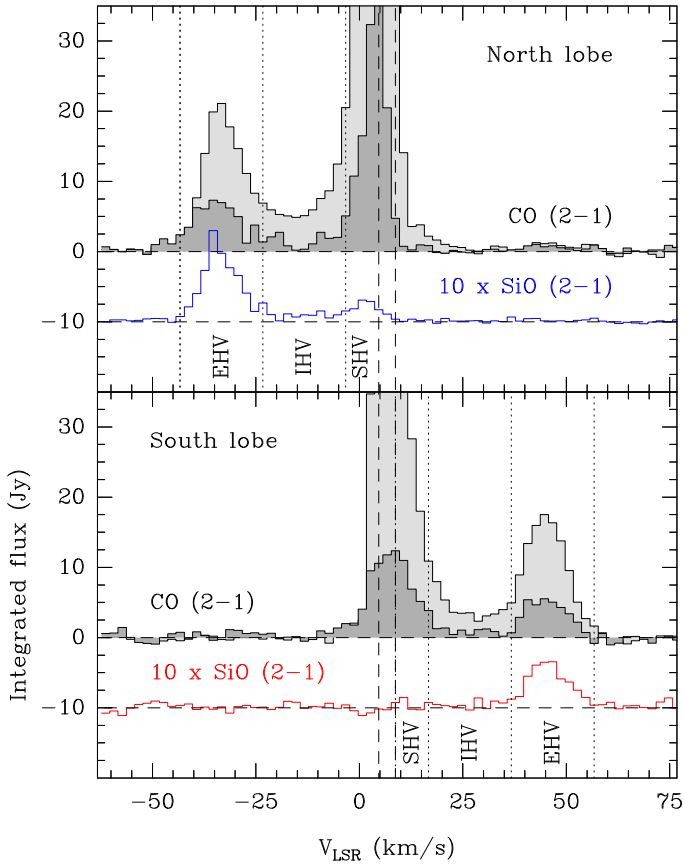}}
\caption{CO(2--1) and SiO(2--1) spectra integrated over the north
(blue) and south (red) lobes of the I04166 outflow illustrating the distinct
appearance of the EHV, IHV, and SHV regimes (see text). Note how the EHV 
gas forms well-defined secondary maxima shifted by about 40 km 
s$^{-1}$ to the blue and red of the ambient component (centered at 
$V_{\mathrm LSR}=6.7$~km s$^{-1}$). The brighter CO(2--1) spectra
in the plot result from combining  interferometer and single-dish data, 
while the weaker counterparts correspond to interferometer-only data.
The SiO(2--1) spectra have been scaled and shifted downwards for clarity.
\label{fig_pdbi_spec}}
\end{figure}

Figure~\ref{fig_cont}
shows maps of the 3.5 and 1.3~mm continuum emission towards the vicinity
of I04166. In both panels, the emission presents a well-defined single peak
near the nominal IRAS position, and no other mm peak is detected
in the mapped region. We interpret this mm peak as the counterpart of the
IRAS source, and we determine its location by fitting the visibilities
at both wavelengths. The results of
these fits, which agree better than one arcsec, indicate that the
source is located at $\alpha(J2000)=4^h19^m42\fs5,$
$\delta(J2000)=+27^\circ13'36''$.

The analysis of the interferometer visibilities also
shows that the mm emission from the central peak 
is not resolved by the observations.
This means that the region responsible for this emission 
must be smaller than our beam, which for the 1.3~mm observations 
is of $1\farcs4$, or
about 200~AU at the distance of Taurus. 
Given this small size, the dust is most likely located in 
a disk around the protostar, although a contribution from the inner 
protostellar envelope
cannot be totally ruled out \citep{jor07,chi08}. For disk
emission in Taurus, it is possible to obtain a reasonably
accurate estimate of the disk mass by assuming that the emitting
dust is approximately isothermal at a temperature of 20~K
\citep{and07}. According to the disk model of
\citet{dal98}, this temperature is reached
in the disk mid plane within the inner 50~AU from 
the central object, so the 20~K radius lies 
well inside our interferometer
beam. Assuming a 20~K temperature and a 
1.3~mm dust emissivity of 0.01~cm$^2$ g$^{-1}$ \citep{oss94},
our measured 1.3~mm flux of $59 \pm 2$~mJy implies a disk 
(gas + dust) mass of approximately 0.02~M$_\odot$.

Information on the dust physical properties can be derived from the frequency
dependence of the dust emissivity. Assuming optically
thin emission from 20~K dust and using the above 1.3~mm flux together with the
measured 3.5~mm flux of $11\pm 0.5$ mJy, our observations imply a 
power-law index for the dust emissivity of $\beta = -0.1 \pm 0.1$
($\kappa_\nu \sim \nu^\beta$). Such a value of $\beta$ is much lower 
than the canonical ISM value of 2  \citep{dra84}, 
but it is similar to the values
found in other YSOs in Taurus, and it can be understood as resulting
from grain growth at the high densities expected in the disk interior
\citep{bec91}.

\section{Overall outflow morphology}

\begin{figure*}
\centering
\resizebox{\hsize}{!}{\includegraphics{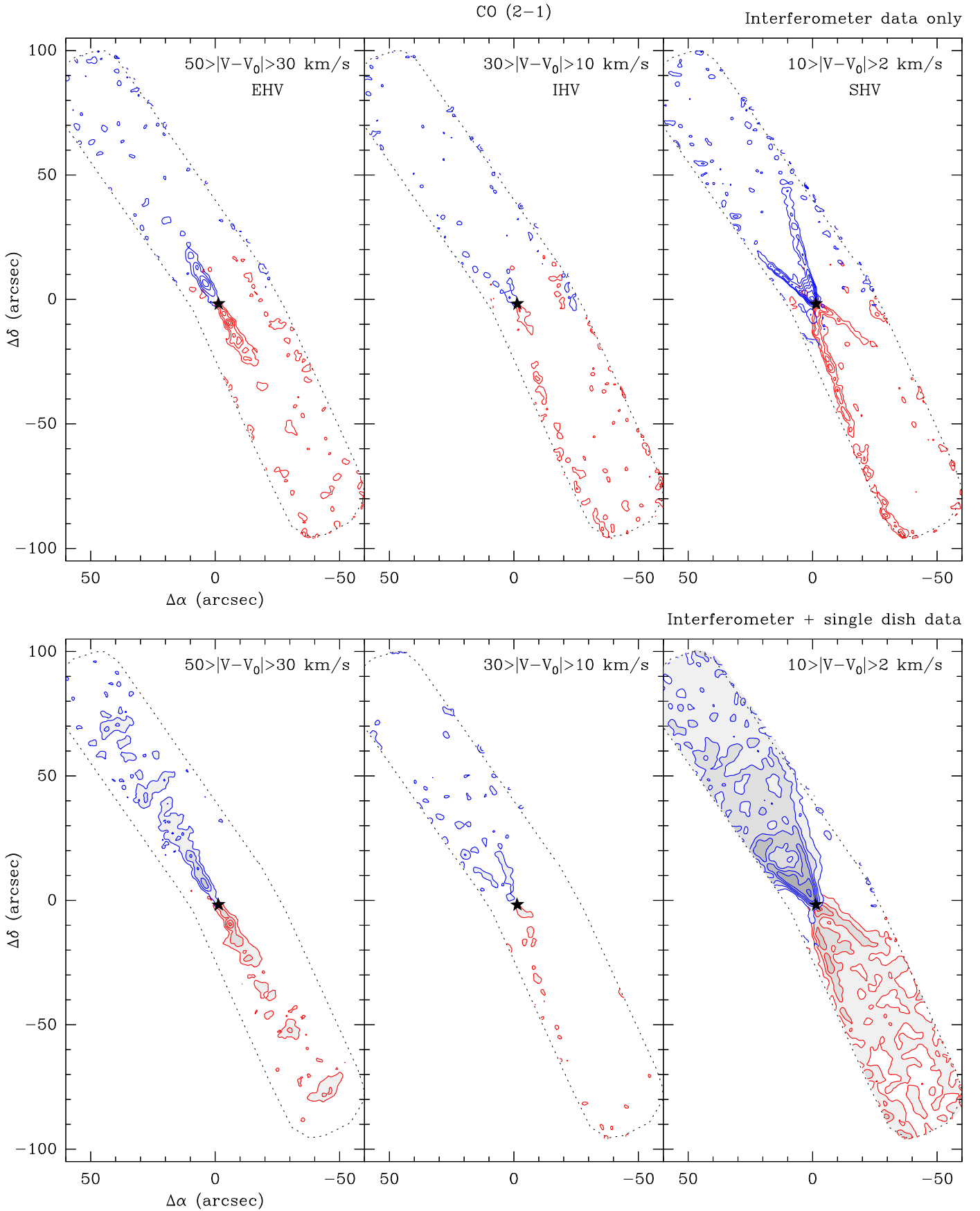}}
\caption{Maps of the CO(2--1) emission towards I04166
integrated over three
velocity ranges referred to, by decreasing velocity,
as extremely high velocity (EHV), intermediate
high velocity (IHV), and standard high velocity (SHV). The top
panels show maps made using interferometer data only, which
enhances the small-scale structure of the emission. The bottom panels
show maps made combining the interferometer data with
visibilities derived from single dish (IRAM 30m) observations,
and are more sensitive to the low-level, extended emission. Note
how the EHV emission consists of a symmetric collection of
highly aligned discrete peaks, while the SHV emission forms
two shells that surround the EHV jet. The offsets are
referred to the phase center of the interferometer observations
($\alpha_{J2000}=4^{\rm h}19^{\rm m}42\fs6$,
$\delta_{J2000}=27\degr13'38''$),
the star symbol indicates the location of the mm continuum peak,
and V$_0$ ($=6.7$~km~s$^{-1}$) corresponds to the velocity of the ambient
core as determined from NH$_3$ observations.
In the interferometer-only maps, the
first contour and interval are at 6~K~km~s$^{-1}$
for the EHV and IHV ranges and 5~K~km~s$^{-1}$
for SHV. In the combined maps, the first contour and interval are
5~K~km~s$^{-1}$ for all the velocity ranges in the red-shifted gas, and
6~K~km~s$^{-1}$ (EHV and IHV) and 7~K~km~s$^{-1}$ (SHV) in the
blue-shifted gas.
The dotted line indicates the region covered by the interferometer
observations. Beam sizes are approximately $2\farcs5\times2''$ in
the interferometer-only maps and $3''\times3''$ in the combined maps.
\label{fig_co_3vel}}
\end{figure*}

\begin{figure*}
\centering
\resizebox{\hsize}{!}{\includegraphics{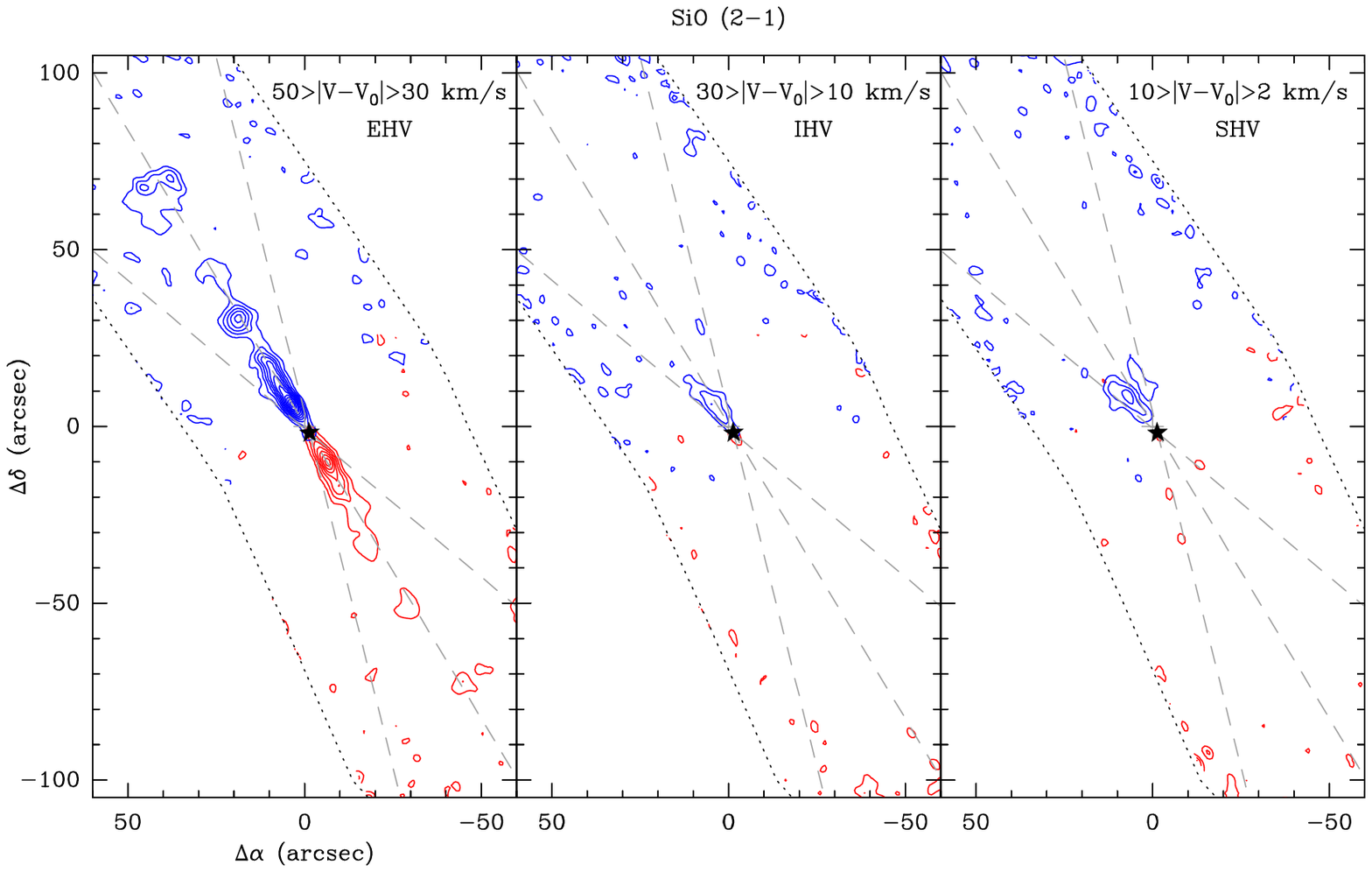}}
\caption{Maps of the SiO(2--1) emission towards I04166
integrated over the same velocity ranges as in Fig.~\ref{fig_co_3vel}.
The long-dashed lines
that converge into the star symbol (mm peak position) represent
the directions of the jet and shells seen in the CO(2--1) map of
Fig.~\ref{fig_co_3vel}, and are presented to aid the eye when comparing the
emission from the two species. Note how the EHV SiO(2--1)
emission presents a similar distribution to the CO(2--1)
EHV emission. The IHV and SHV SiO(2--1) emission
is only detected toward the northern blue lobe
and seems associated to the CO shells.
First contour and interval are 2~K~km~s$^{-1}$ and
1.5~K~km~s$^{-1}$ for the EHV and IHV ranges of the
blue and red emission, respectively, and 1.5~K~km~s$^{-1}$
and 1~K~km~s$^{-1}$ for the SHV range of the
blue and red emission, respectively.
Offsets, star symbol, and dotted lines as in Fig.~\ref{fig_co_3vel}
Only interferometer data were used to generate the maps,
and the synthesized beam FWHM is approximately $4''\times 3''$.
\label{fig_sio_3vel}}
\end{figure*}

To study the properties of the 
I04166 outflow, we now turn to our interferometer observations
of the CO(2--1) and SiO(2--1) emission.
From single dish data, TSJB04 found that the CO(2--1) outflow 
emission presents three distinct velocity regimes referred to as
extremely high velocity (EHV), intermediate high velocity (IHV), and
standard high velocity (SHV), and which correspond to
velocity displacements from the ambient cloud of 50 to 30 km s$^{-1}$
(EHV), 30 to 10 km s$^{-1}$ (IHV), and 10 to 2 km s$^{-1}$ (SHV).
This division of the emission 
in three velocity regimes is also apparent in our
high resolution data, and is illustrated 
in Fig.~\ref{fig_pdbi_spec}
using the CO(2--1) and SiO(2--1) spectra integrated over each outflow 
lobe (for CO, both interferometer-only and
combined single-dish and interferometer data are presented).
As Fig.~\ref{fig_pdbi_spec} shows, the EHV regime appears in the spectra as a pair of
secondary peaks symmetrically shifted about 40 km s$^{-1}$ to the red and blue
of the ambient cloud (at $V_{\mathrm LSR}=6.7$~km~s$^{-1}$,
see TSJB04), while the SHV regime appears as the characteristic high velocity 
red and blue wings of an outflow. The IHV regime is characterized as
the transition between the two other regimes, and its intensity 
is significantly weaker than that of the EHV and SHV gas.
As the figure also shows, the three regimes are well detected in CO(2--1), 
while most of the SiO(2--1) emission belongs to the EHV regime.

The spatial location of the three outflow velocity regimes in CO(2--1) is
presented in Fig.~\ref{fig_co_3vel} with two sets of maps.
The top set shows interferometer data only, so the maps tend
to magnify the highest spatial frequencies and therefore stress
the most compact structures. The bottom set of maps
was made adding single dish data to the interferometer
visibilities and therefore contains the more diffuse emission.
As can be seen, both sets of maps reveal the same
general behavior: (i) the emission is bipolar with respect to the
millimeter source, (ii) it presents a very high degree of collimation 
at the highest velocities (EHV, leftmost panels), and (iii) appears cometary
in the lowest velocity range (SHV, rightmost panels). The main differences
between the two sets of maps are the additional emission
seen at large distances from the protostar in the combined EHV
map and the diffuse component filling the outflow
lobes that appears in the combined SHV map. In
both sets of maps, the emission at intermediate velocities (IHV, middle
panels) is significantly weaker than in the other two velocities
and is qualitatively similar to the emission of the slowest gas.

The most remarkable feature of the maps in Fig.~\ref{fig_co_3vel} is the
very different spatial distribution of the EHV and SHV regimes.
The high collimation of the EHV emission, especially close to the
millimeter source, where it remains unresolved by our $\approx 2\farcs5$
(350 AU) beam, indicates that the fastest gas in the outflow must be located 
in a jet-like component that travels along a straight line through
the center of each lobe.
The cometary shape of the SHV emission, on the other hand, 
suggest that the lowest velocity gas in the outflow moves along two
almost-conical shells that surround symmetrically the 
high velocity jets and have the IRAS source at their vertex. 
This very different geometry of the extreme velocity regimes
suggests that each outflow lobe consists of two distinct 
physical components, a jet and a shell, and that there is
little or no connection between the two. 
The data, in particular, seem inconsistent with
an outflow distribution where all the gas is located in two shells,
and where the difference between the jet and the shell arises only
from projection, with the shell being the part of the flow moving 
orthogonal to the line of sight and the jet being the most blue or
red shifted part of the same conical flow (see \citealt{taf97} for an
application of such a model to the Mon R2 outflow). 
Such a geometrical interpretation of the emission
would require that there is a series of 
maps at intermediate velocities 
where the intensity is similar and the outflow shells 
converge continuously toward the center of the lobes. As seen in
the integrated maps of Fig.~\ref{fig_co_3vel} such a trend is not found, and as
seen in the spectra of Fig.~\ref{fig_pdbi_spec}, the intermediate velocity regime
is much weaker than the EHV emission, which clearly 
is not a continuation of the outflow wing.
The shell-only model, in addition, predicts that two jet features
should be seen per lobe, one arising from the front and the other from the
back of the shell, and that the velocities of these two jet features should be
symmetric with respect to the shell regime.
Given that in I04166 the difference in velocity between the jet and 
the shell is larger than 30 km s$^{-1}$, and that the
shell is moving about 10 km s$^{-1}$ with respect to the ambient gas,
the shell-only model predicts that each outflow lobe should have a blue 
and a red jet, which is clearly not the case.
We therefore conclude that the shells and jets in I04166 
are not the result from a projection effect, but  that they represent 
two separate components of the bipolar outflow.
In the next three sections we discuss with more
detail the nature and characteristics of these two components,
together with some implications for the underlying physics of the
driving wind. In the remainder of this section we continue with 
a description of the overall properties of the outflow.

A complementary view of the I04166 outflow comes from the SiO(2--1)
emission, which was observed simultaneously with CO(2--1)
and has a factor-of-two lower resolution. The maps of 
this emission,
integrated in the same velocity ranges used for CO, are presented
in Fig.~\ref{fig_sio_3vel}. These interferometer-only data recover most of the 
flux, and
therefore reflect the true distribution of SiO emission 
in the region.  As Fig.~\ref{fig_sio_3vel}
shows, the SiO(2--1) emission in the EHV range has a 
jet-like distribution very similar to that seen in CO(2--1), 
especially when taking into account the lower
angular resolution of the SiO data. As we will show in section 6, 
the CO and
SiO emissions in this velocity range agree with each other both 
in spatial distribution
and kinematic structure, indicating that they originate from the 
same jet-like component of the outflow. The intermediate and low 
velocity ranges (center and right panels in Fig.~\ref{fig_sio_3vel}) show much 
weaker SiO(2--1). The emission is only detected towards
the northern blue lobe, and it has a distribution that resembles
the CO blue shell. This asymmetry in the SiO emission is real,
and it is probably related to the strong (factor of 3) asymmetry
found in the intensity of the CO emission for the same velocity 
range (TSJB04).

\section{The shells}

\begin{figure}
\centering
\resizebox{7cm}{!}{\includegraphics[angle=270]{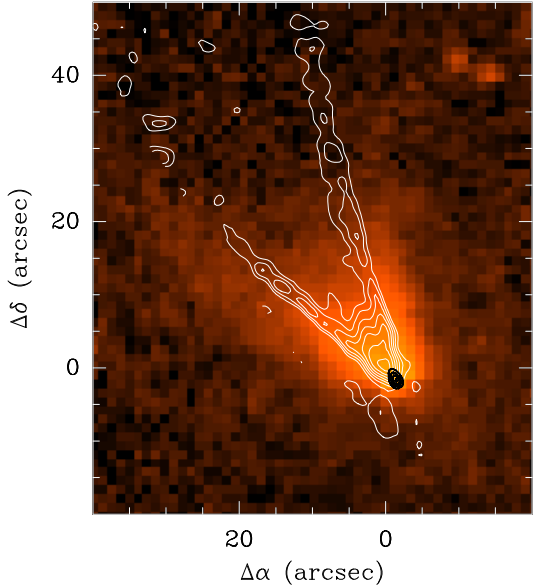}}
\caption{Superposition of the CO(2--1) SHV interferometer-only 
map (white contours) and the
archival $3.6~\mu$m Spitzer IRAC image of I04166 (red scale)
illustrating the coincidence of the outflow blue shell and the 
NIR scattering nebula. The black contours indicate the 1.3~mm
continuum emission from the central protostar.
Offsets origin and CO contours as in Fig.~\ref{fig_co_3vel}.
\label{fig_co_shell}}
\end{figure}

As discussed in the previous section, the distribution of CO(2--1)
emission in the SHV range suggests that the slowest
molecular outflow is confined to the walls of two opposed conical
shells. Such a distribution has been seen in other outflows
by a number of authors
\citep[e.g.,][]{mor87,mey91,bac95,arc06,jor07}
and is naturally explained if the shells represent the walls 
of two cavities that have been evacuated by the outflow.
In the case of I04166, such an
interpretation is strengthened by the recent
(publicly-available) NIR images of the region taken with the Spitzer Space 
Observatory's IRAC camera as part of GO program 3584 (PI: D. Padgett), 
which show a cometary scattering nebula
extending north of the mm source. As illustrated in Fig.~\ref{fig_co_shell}, 
this NIR nebula 
matches nicely 
the blue CO emission in both direction, shape, and size, as it is expected 
if the outflow lobe has evacuated the cavity responsible for the nebula. 
Detailed modeling of the NIR emission shows that the Spitzer 
images can be reproduced with a scattering nebula that has the geometry 
suggested by CO if this emission delineates the boundary of the cavity
(A. Crapsi, private communication). Furthermore, the
opening angle we measure from the CO data ($\approx 32^\circ$) is in perfect
agreement with the angle determined independently by \citet{sea08}
from the NIR Spitzer images (note how in Fig.~\ref{fig_co_shell} 
the NIR emission penetrates slightly
the walls of the CO cavity, as expected from the scattering model, but it does so in a
way that preserves the opening angle).  Finally, we note that
the recent N$_2$H$^+$(1--0) OVRO
map from \citet{che07} reveals an hourglass distribution of the
dense gas that has its waist perpendicular to the outflow axis. This is 
the geometry expected if the outflow had carved a pair
of opposed cavities in the dense core that surrounds the protostar.

We can use the CO and NIR data to constrain the orientation of the
outflow axis with respect to the line of sight
assuming the shells are conical. Both the absence 
of direct light from the central protostar in the optical/NIR images 
and the lack of overlap between the north and south CO lobes 
imply that the angle between the outflow direction and the 
line of sight is larger than half the cavity opening angle,
or $\approx 16^\circ$. On the other hand, the lack
of a scattering nebula towards the southern red lobe 
and the lack of mixing between blue and red gas
in each of the outflow lobes 
(assuming the gas moves along the shells, see below) 
imply that the orientation
angle should be smaller than 74 (=90-16) degrees
\citep{cab90}.
Although any inclination between the above two extreme limits could be
a priori consistent with the observations, the true value is
likely to be rather intermediate. A detailed modelling of the
NIR scattering nebula (A. Crapsi, private communication)
suggests an inclination of 45 degrees,
and in the following, we will use
this angle as a reference for any kinematics estimate.

An inspection of the individual channel maps in the low
velocity regime
shows that the opening angle of the CO shells does not
change with velocity over the whole SHV range.
This behavior suggests that the outflow motions we observe
are mostly directed along the shells, and that any perpendicular
component of the velocity field 
is likely to be negligible (see \citealt{mey91} for
a detailed analysis of a similar case in Mon R2). 
The CO shells, in addition, can be seen in velocity maps 
that span a range of at least 10 km s$^{-1}$, so 
the material in the SHV regime must span a similar
range of longitudinal velocities. This coexistence of 
material moving along the shells with a large range of 
velocities can be understood if
the low velocity outflow consists of a shear
flow of accelerated ambient gas moving 
along the walls of the evacuated
cavities. To produce such a velocity pattern, the 
driving wind of the outflow must have a speed of
at least 10 km s$^{-1}$ with respect to the ambient cloud.
Although the EHV gas satisfies this requirement
and has enough momentum (TSJB04),
it is unlikely to constitute the accelerating agent of
the SHV gas. As we will see in the next section, the EHV
component shows no evidence for momentum transfer to
the ambient gas and its opening angle is smaller
than the 
$>30^\circ$ opening angle measured in the SHV shells. 
The SHV gas, therefore, is most likely accelerated by a
different (and invisible) component that emerges
from the central YSO with a velocity of at least 10 km s$^{-1}$
and an opening angle of at least $30^\circ$. The straight cavity
walls and the highly directed
velocity field of the CO SHV suggests that this component has
certain degree of collimation, although a more detailed
analysis of the interaction between the outflow and the dense
gas is needed to reach a firm conclusion. On-going PdBI 
observations of the core material around I04166 will hopefully
shed new light on this important issue.

\section{The jets}

\subsection{Integrated emission}

\begin{figure}
\centering
\resizebox{\hsize}{!}{\includegraphics[angle=270]{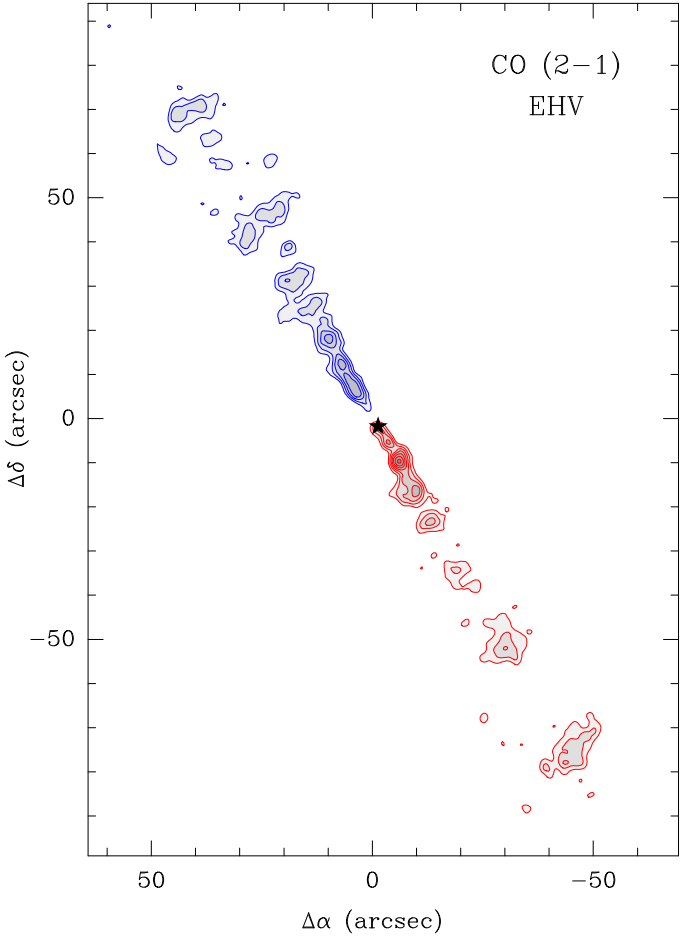}}
\caption{Clipped CO(2--1) map of the EHV regime showing an enhanced
view of the fastest part in the I04166 outflow. The map has been
made using single dish and interferometer visibilities,
and before adding the individual channels, the emission
has been clipped at 0.5~K (blue) and 0.3~K (red).
First contour and interval are 3.5~K~km~s$^{-1}$ (blue) and 
2.5~K~km~s$^{-1}$ (red). Offsets and star symbol are as in Fig.~\ref{fig_co_3vel}.
\label{fig_clip}}
\end{figure}

The maps of integrated intensity for the EHV range shown in Fig.~\ref{fig_co_3vel}
do not provide the best view of the highest velocity component in the
outflow, as the 20 km s$^{-1}$-wide velocity integral dilutes
significantly the emission and makes some low-level features
disappear in the noise of the emission-free channels. Due to the internal velocity
structure of the EHV gas (see next section), maps
with a narrower velocity range do not provide a complete view
of this component either, so to achieve the best representation of the EHV gas 
we resort to maps where the low-level emission has been clipped
before integrating in velocity. Figure~\ref{fig_clip} presents the clipped
version of the combined interferometer plus single dish CO(2--1)
emission map and shows a better defined structure than the equivalent
non clipped version of Figure~\ref{fig_co_3vel}.
As the clipped map illustrates, the EHV gas, in addition of
being highly collimated, is very fragmented, and it seems to
consist more of a collection of discrete emission peaks than
a smooth and continuous jet. 

\begin{table}
\caption[]{Properties of EHV peaks from 2D gaussian fits.
\label{tbl-1}}
\[
\begin{array}{ccccc}
\hline
\noalign{\smallskip}
\mbox{Name} & \mbox{Position$^{(1)}$} & \mbox{I[CO(2--1)]$^{(2)}$}  & 
\mbox{I[SiO(2--1)]$^{(2)}$} & \mbox{M$^{(3)}$} \\
& ('','') & \mbox{(K km s$^{-1})$} & \mbox{(K km s$^{-1})$} & \mbox{($10^{-5}$ M$_\odot$)}  \\
\noalign{\smallskip}
\hline
\noalign{\smallskip}
\mbox{R1} & (-2.3,-3.2) & 11.7 & 7.2 & 2.2 \\
\mbox{R2} & (-5.0,-7.6) & 20.4 & 11.9 & 4.2 \\
\mbox{R3} & (-7.5,-14.1) & 12.9 & 6.9 & 5.1 \\
\mbox{R4} & (-11.9,-21.3) & 9.6 & 2.4 & 2.4 \\
\mbox{R5} & (-18.4,-33.1) & 4.8 & 1.5 & 3.4 \\
\mbox{R6} & (-28.8,-48.9) & 7.5 & 2.0 & 5.9 \\
\mbox{R7} & (-43.8,-73.0) & 8.1 & 1.3 & 7.4 \\
\mbox{B2} & (6.0,9.7) & 19.2 & 18.9 & 5.9 \\
\mbox{B3} & (8.3,13.8) & 17.1 & 20.3 & 2.3 \\
\mbox{B4} & (11.2,19.8) & 15.3 & 13.1 & 3.1 \\
\mbox{B5} & (18.8,33.1) & 10.2 & 9.0 & 6.0 \\
\mbox{B6} & (26.5,46.5) & 9.0 & 2.0 & 8.8 \\
\mbox{B7} & (42.2,71.5) & 7.2 & 5.4 & 8.0 \\
\hline
\end{array}
\]
\begin{list}{}{}
\item[Notes:] (1) Offsets with respect to the mm-continuum peak at 
$\alpha(J2000)=4^h19^m42\fs5,$ $\delta(J2000)=+27^\circ13'36''$;
(2) peak intensity from 2D gaussian fit; (3) see text for assumptions.
\end{list}
\end{table}

\begin{figure*}
\centering
\resizebox{18cm}{!}{\includegraphics[angle=-90]{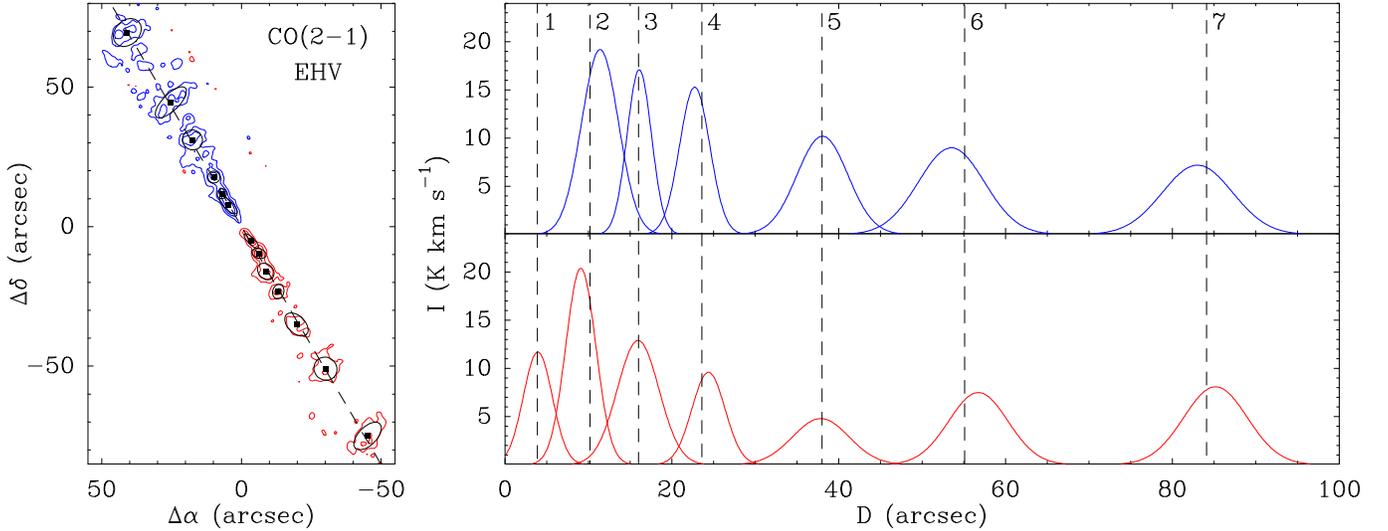}}
\caption{{\em Left:\ } EHV CO(2--1) clipped emission map as in Fig.~\ref{fig_clip} 
(blue and
red contours) with the results from 2D gaussian fits to the emission
peaks (in black). The solid squares represent the centers of the gaussians and the
ellipses represent the fit half maximum contours. The dashed line
traces the result
from a least squares fit to the gaussian centers and has a dispersion
of less than one degree (see text). {\em Right:\ } 2D gaussian fit results as
a function of distance to the mm peak (blue lobe in top panel and red lobe
in bottom panel). For each EHV peak, the plot shows a gaussian
having the fit intensity, distance to the mm peak, and
a width given by the geometric mean of the widths from the
2D fit. Note the blue/red symmetry in the position of the peaks,
and the decrease in intensity and widening of the gaussians
with distance to the source.
\label{fig_gauss}}
\end{figure*}

To study the properties of the EHV peaks,
we have fitted their CO(2--1) emission in the clipped map 
with 2D gaussians using the GAUSS\_2D
routine of the GILDAS program GreG.
The results of this fitting are summarized in Table 1 and 
presented graphically in the left panel of Fig.\ref{fig_gauss} superposed to the
EHV emission. With these fits, we can quantify some of the properties
of the peaks, and in particular, study the symmetry of their distribution.
This is presented graphically in the right panel of
Fig.~\ref{fig_gauss}, where we have plotted the profiles of the gaussian fits as a function of 
distance to the mm source. As can be seen, except for peak 1, which seems missing 
in the blue lobe (a possible result of the clipping), 
all the other 6 peaks appear
both in the red and blue outflow lobes with an almost perfect
one-to-one correspondence. In fact, 
the mean difference between the distance of the blue and red peaks
to the central source
is less than $2''$, which is comparable to the beam 
size of the observations.
Such a regular distribution of the peaks with respect to the central
source makes I04166 one of the most symmetric outflows known.

The 2D fits to the CO emission provide additional information on the EHV component.
A least squares fit to the peak positions
(dashed line in the left panel of Fig.~\ref{fig_gauss}) indicates that that the 
outflow
lies at PA = $30\fdg4$ (east or north) with a dispersion of $0\fdg2$.
Such a low angular dispersion confirms the 
almost perfect alignment between the CO peaks,
and suggests that any precession or bending 
of the jet-like component of the 
outflow must be significantly smaller than one degree. 

While the peak alignment is better than one degree, the
opening angle of the EHV jet is larger. As can be
seen even in the non clipped images, the EHV
peaks broaden with distance from the mm source,
especially in the direction perpendicular to the jet. This
broadening of the peaks is accompanied in some cases 
(like B6, B7, and R7) by a slight curvature of
the emission, which gives the peaks
the appearance of bow-shocks propagating away from the
source. Using the results from the 2D fits
to the 3 outermost peaks of each lobe (numbers
5 to 7) we derive from a least squares fit an opening
angle of 10 degrees for the EHV gas. This angle is 
3 times smaller than the opening angle of the shells,
suggesting again that the jet and the shells are 
independent outflow components. In particular, it seems unlikely that 
in the region we have observed,
the shells have been accelerated by the precession or broadening 
of the jet.

In addition to broadening, the peak CO emission from the EHV 
features weakens with distance to the mm source, as can
be seen in the right panel of Fig.~\ref{fig_gauss}. This effect, which 
is again symmetric in the two lobes,
is more than compensated by the broadening of the features,
so the outermost EHV peaks have a slightly (factor of 2)
larger integrated intensity than the innermost peaks. 
We can use this CO(2--1) integrated intensity to
estimate the mass of each peak. The detection of SiO(2--1)
guarantees a relatively high density (section 6.3),
so LTE conditions are likely to apply to the CO-emitting gas.
A more complex issue concerns the temperature
and CO abundance of this gas. In TSJB04, we
assumed a $T_{\mathrm ex}$ of 20~K and a standard 
Taurus CO abundance of $8.5\; 10^{-5}$ \citep{fre82}.
As discussed below, however,
the kinetic temperature of the EHV gas is likely to be
significantly higher than 20~K, and the CO abundance
is also expected to exceed $8.5\; 10^{-5}$ if this gas
represents a protostellar wind instead of
accelerated ambient material. Fortunately, the mass
determination depends on a ratio where the two effects
almost cancel each other, so if, for example, the excitation 
temperature is as high
as 500~K (section 6.3) and the CO abundance is as high as
$4\; 10^{-4}$ (predicted by the protostellar wind models
from \citealt{gla91} and discussed also in section 6.3), the
mass estimate is only a factor of 2 higher than predicted 
using the assumptions in TSJB04. Thus, and
for consistency with the interpretation in section 6.3, we
use the higher temperature and higher abundance assumptions,
although the effect of this choice is relatively 
small given all the uncertainties in the estimate. Using these
values, we have estimated the masses given in Table~1, 
which indicate that the 
typical mass of an EHV peak is 
$5\; 10^{-5}$~M$_\odot$ 
and the total mass 
in this outflow component is about $6.5\; 10^{-4}$~M$_\odot$. Dividing this
mass by the kinematical age of the outermost 
peak (1,400 yr, assuming a velocity of
40 km s$^{-1}$), we derive a mean mass loss for the EHV gas of
about $5\; 10^{-7}$~M$_\odot$~yr$^{-1}$.

The combination of (i) symmetry in the location of the peaks with
respect to the mm source, (ii) high collimation, and (iii)
systematic broadening with distance makes the EHV 
molecular emission of I04166 very similar to the optical
emission of some highly collimated Herbig Haro jets like HH34 and HH111
\citep{rei01}. Molecular outflows from other Class 0 sources like L1448-mm
\citep{bac90},
HH211 \citep{gue99}, and HH212 \citep{cod07,cab07,lee08}, among others, show similar
(although less extreme) behavior, indicating that the properties of
I04166 are not peculiar to this source but 
reflect properties of a wide class of bipolar outflows. 
The symmetry in the
location of the peaks at each side of the central source, in particular,
suggests that the production of the EHV peaks is related to 
some type of episodic event in the outflow source,
which based on the kinematic ages of the peaks, has a time scale of
the order of 100 years. In addition, 
the high collimation of the emission indicates that the source is able to 
produce a jet-like component simultaneously with 
the wider wind responsible for the shells.
To further constrain the origin of the EHV gas, we
need to study its kinematic properties.

\subsection{Kinematics}

\begin{figure*}
\centering
\resizebox{\hsize}{!}{\includegraphics[angle=270]{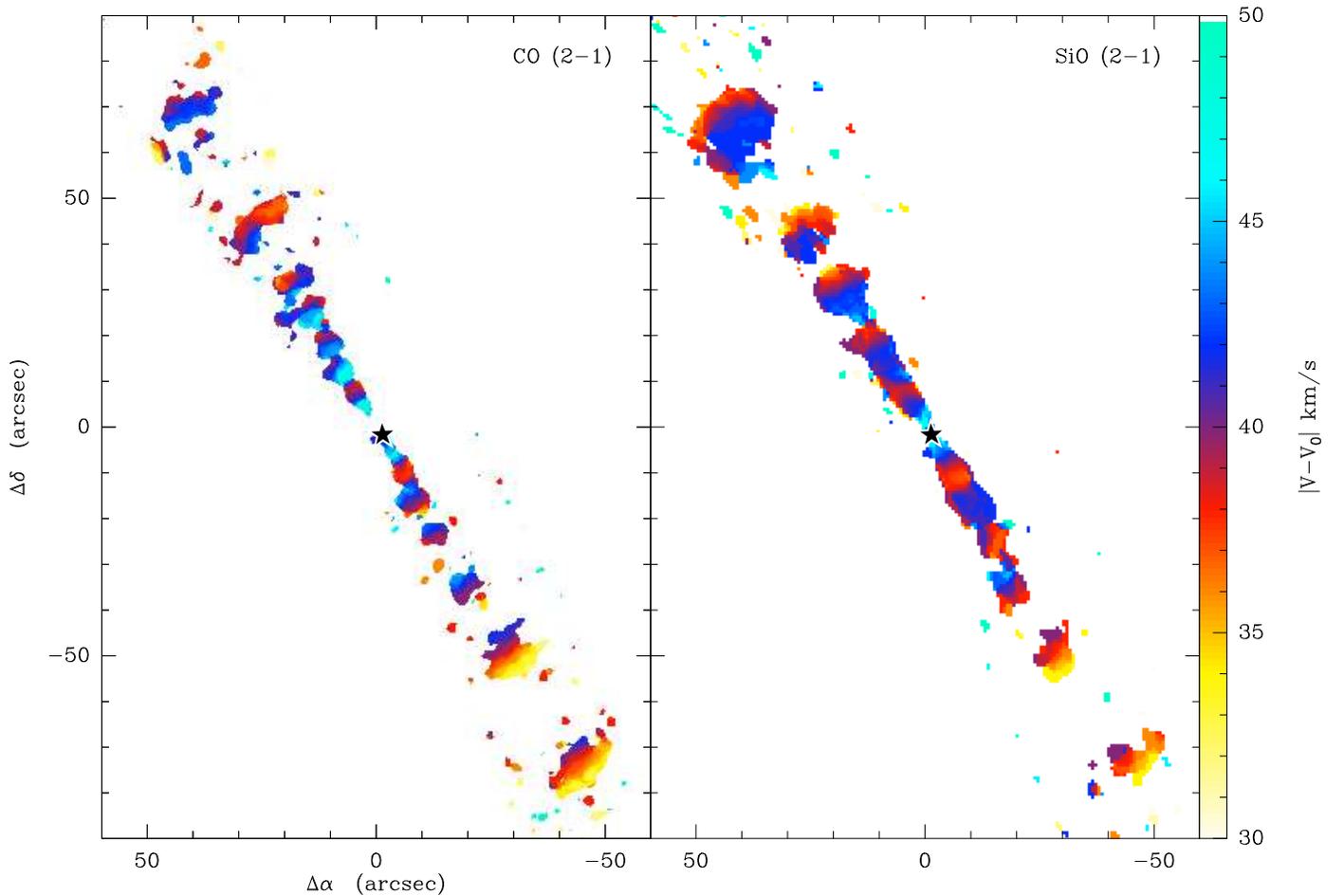}}
\caption{CO(2--1) (left) and SiO(2--1) (right) first momentum maps
showing the difference between the velocity of the EHV outflow 
gas and the ambient cloud in absolute value. Note the 
oscillating velocity pattern
To enhance the S/N, the emission has been clipped before
making the EHV maps.
Offset center and star symbol are as in Fig.~\ref{fig_co_3vel}.
\label{fig_zero_mom}}
\end{figure*}

\begin{figure*}
\centering
\resizebox{12cm}{!}{\includegraphics[angle=270]{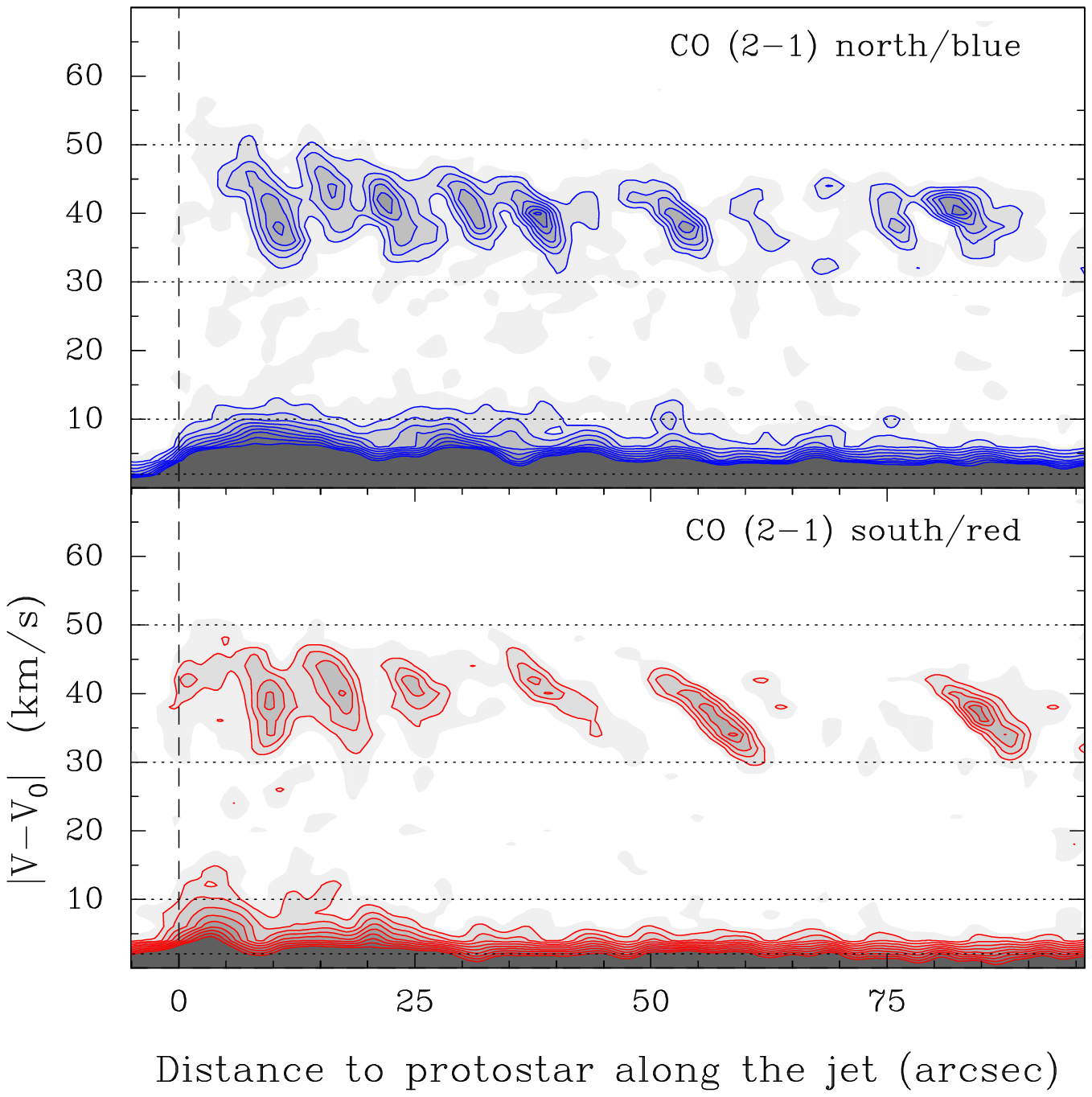}}
\caption{Position-velocity diagrams of the CO(2--1) emission along the
outflow axis (blue gas in top panel and red gas in bottom panel).
For each lobe, the absolute value of the difference between the outflow
and the ambient cloud velocities is presented. Note the discrete EHV peaks
in the $30<|V-V_0|<50$~km~s$^{-1}$ interval (enclosed by 
dotted lines). Each of these peaks
presents an internal velocity gradient with the gas closer to 
the protostar moving faster than the gas at further distances. 
Despite these gradients, the mean speed of the EHV gas remains
constant at about 40~km~s$^{-1}$, and the overall velocity profile has a 
characteristic sawtooth pattern. First contour and interval are
0.4~K and 0.2~K respectively, and $V_0=6.7$~km~s$^{-1}$. To enhance
the signal-to-noise ratio, the emission has been convolved by $5''$
perpendicular to the direction of the cut. No clipping has been applied
to these data.
\label{fig_pv}}
\end{figure*}

In addition to high collimation and symmetry, the 
EHV peaks in I04166 present a systematic pattern of internal velocity
gradients. 
Figure~\ref{fig_zero_mom} illustrates this pattern with the first moment of 
absolute velocity measured with respect to the ambient cloud 
($V_{\mathrm LSR}=6.7$ km s$^{-1}$, see TSJB04). As it can be seen, both 
in CO(2--1) and SiO(2--1) the emission alternates 
between fast and slow over the whole length of the blue and red jets.
A more quantitative view of the pattern is provided
by the position-velocity (PV)
diagrams along the outflow axis shown in Fig.~\ref{fig_pv} (again, the velocity 
is measured with respect to the ambient cloud). These diagrams 
illustrate how each EHV peak presents a velocity structure in which
the gas closer to the mm source (the ``tail") moves faster
than the gas further away from the mm source (the ``head").
The difference between the head and tail velocities 
is 10-15 km s$^{-1}$, and the change between the two values is
almost linear with distance along the jet. 
As the figure shows, the length of the EHV peaks along the
outflow axis increases systematically
with distance to the source, and this
makes the slope of the velocity gradients change from rather
steep near the outflow origin to flatter at large
distances. A simple (pencil and ruler) fit to the data
indicates velocity gradients larger than 2 km s$^{-1}$ (100~AU)$^{-1}$
for the three inner peaks and values slightly
lower than 1  km s$^{-1}$ (100~AU)$^{-1}$ for the three outer
peaks. 

As the PV diagram shows, the gradients in the EHV gas are
only local. Each gradient affects the internal
velocity of one EHV peak, but it does not propagate downstream 
or upstream to the neighboring peaks.
All EHV peaks, therefore, have similar head and tail velocities,
and this behavior gives
the velocity field a characteristic sawtooth pattern along the
jet axis. As a result, the EHV gas keeps an almost constant 
mean velocity of about 40 km s$^{-1}$ despite the 
steep gradients inside each of the peaks.
Such a velocity pattern can hardly be explained if the gradients
arise from an interaction
between the outflow and the ambient gas, as this would require a 
steady deceleration of the gas along the jet due to
the gradual transfer of momentum to the ambient material.
To keep a constant mean velocity, the bulk of the gas has to be moving
without external perturbations,
from the very vicinity of the mm source to the
furthermost region mapped by our observations. This behavior seems
incompatible with an interpretation in which the EHV
gas represents ambient material that has been accelerated by the outflow
either through entrainement or bow shocks.  

A more likely interpretation of the observed 
velocity field is that the EHV gas represents material
directly ejected by the star/disk system, or loaded
into the outflow in the very vicinity ($<100$~AU) of the mm source, 
and that is traveling in a straight line without much interaction
with the surrounding cloud.
The pattern of a fast tail and a slow head in each EHV peak, however, 
indicates that this gas does not move like  
a collection of bullets, as in that case a signature of constant velocity
would be expected inside each peak. The pattern also rules out that the gas
moves as flying shrapnel ejected in 
a series of discrete explosions, because
in that case the material inside each EHV peak would have sorted itself in velocity,
with the faster gas having traveled further and therefore lying ahead,
which is the opposite to what is observed.
The sawtooth velocity pattern, on the other hand, seems consistent with 
the motion predicted for a pulsed jet with internal working surfaces.
This type of model was initially
proposed by \citet{rag90} to explain some of the observed properties
of HH objects,
and it has subsequently been the subject of extensive analytic and numerical
work \cite[e.g.,][]{rag92, har93,sto93,deg94,bir94,mas02}.
In a pulsed jet, supersonic variations in the velocity of
ejection (likely caused by variability in the accretion)
give rise to a train of discrete compressions that
occur when the gas emitted during a period of fast ejection
overtakes slower material emitted before. When this overtaking occurs,
a 2-shock structure is formed, consisting of a forward shock
where the slow jet material is accelerated and a reverse shock where
the fast jet material is decelerated. Such a 2-shock structure is
often referred to as an internal working surface (IWS), and
both simulations and analytic work show that it moves along the jet 
and grows in size as more jet material is incorporated
through the two shocks (see previous references). The material inside each
IWS is highly compressed, and a fraction of it is squirted
laterally into the cocoon, giving rise to bow shaped
structures clearly discernible in the 2D and 3D simulations \citep{sto93}
and similar to those seen in the map of Fig.~\ref{fig_clip}. 

\begin{figure}
\centering
\resizebox{\hsize}{!}{\includegraphics{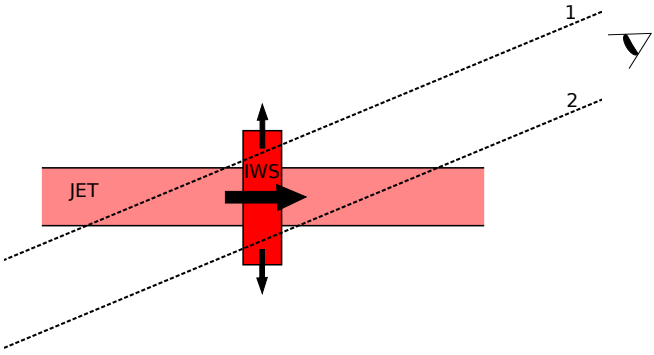}}
\caption{Schematic diagram showing how the sideways ejection
of material in an internal working surface (IWS)
makes the upstream gas appear
faster than the downstream gas. In the upstream
line of sight (labeled 1), the radial components of the
jet motion (rightward arrow) and the sideways ejection
(upward arrow) are both directed towards the observer
and their contributions reinforce.
In the downstream line of sight (labeled 2),
the sideways ejection (downward arrow) contributes with
the opposite sign to the radial velocity.
Note that this effect does not rely on a specific orientation
other than to increase or decrease the strength of the feature (as
opposed to its sense).
\label{fig_cartoon}}
\end{figure}

Both numerical simulations and analytic work predict that
the velocity field of the gas along a pulsed jet should have a
sawtooth profile, and that the IWSs should be located in the 
drop sections of the teeth.  
Although such a profile is similar to the PV diagram of Fig.~\ref{fig_pv}, 
the velocity drops we observe in I04166 probably do not arise from 
velocity gradients {\em along} the jet axis. This is so
because the velocity drop seen in the models 
corresponds to the {\em pre-shock} gas, while the observed CO and SiO
emission most likely represents {\em post-shock} material (see below).
As discussed by \citet{smi97b} 
and seen in the simulations by \citet{sto93}
and \citet{sut97}, the axial velocity of the shocked gas
inside a working surface should in fact increase with distance to
the source, and this is the opposite behavior to what we observe in the PV diagram. 
The compression {\em along} the jet, however, is not the only velocity
gradient present in the gas of the IWS. As mentioned before, part of the 
shocked gas is squirted sideways from the jet, 
and this velocity component needs also to be considered when predicting the
PV diagram of a pulsed jet. Indeed, \citet{sto93} have used a numerical
simulation to predict the PV diagram expected from the observation of such a
jet, and their results show that the lateral ejection of
material dominates over the axial compression in
the observed kinematics of the emitting gas.
The PV diagrams predicted by these authors present striking similarities
with the PV diagrams of Fig.~\ref{fig_pv}, as they
consist of a series of discrete sections, each of them with a fast tail and
a slow head (see Figs. 13 and 16 in \citealt{sto93}). 
The production of the velocity gradient in one of these
sections is illustrated in Fig.~\ref{fig_cartoon}, which shows how in the 
upstream line of sight, the lateral ejection adds an extra component to the radial
projection of the jet velocity, while in the downstream line of sight,
the component is subtracted and the apparent gas velocity is decreased.
Numerical simulations, in addition, predict
PV diagrams with a systematic flattening of the sawtooth pattern 
with distance to the emitting source,
probably resulting from a decrease in the lateral velocity due to the
weakening of the internal shock. This is again in good agreement
with the observations of I04166, and 
although the \citet{sto93} 
model uses physical conditions expected for
an atomic jet (fast, warm, and low density),
the works of \citet{sut97} and \citet{smi97a}  
show that many of characteristics
of the propagation and the kinematics of atomic jets hold for their 
molecular counterparts.

If the teeth in the PV diagram represent
lateral ejection of material from the IWSs, their width in velocity 
should be consistent with the observed broadening angle of the jet,
which is determined by the ratio
between the perpendicular
and parallel components of the true velocity field \citep{lan59}.
From the PV diagram in Fig.~\ref{fig_pv}, we estimate that the 
(projected) mean jet velocity is 40 km s$^{-1}$, and that the
sideways ejection velocity is about 5 km s$^{-1}$ (as each tooth
has a width of about 10 km s$^{-1}$ in radial velocity).
Thus, the ratio between the perpendicular and parallel components 
of the velocity field in the jet should be
$5/40 \cot \theta$, where $\theta$ is the the angle of the 
jet direction with the line of sight. Assuming $\theta=45^\circ$,
the above ratio implies a jet full opening angle of $14^\circ$, which is 
in reasonable agreement with the $\approx 10^\circ$ measured from the
CO(2--1) map in section 6.1.
We thus conclude that the sawtooth pattern 
in the PV diagram corresponds to 
the sideways ejection of the post-shock gas in internal working surfaces.

\subsection{CO versus SiO} 

Figures \ref{fig_co_3vel}, \ref{fig_sio_3vel}, and \ref{fig_zero_mom}
show that the EHV emission of 
CO(2--1) and SiO(2--1) have very similar
spatial distributions and kinematic behavior.
Discounting the factor-of-2 lower angular resolution of the
SiO(2--1) data and the limited S/N of all maps,
the EHV data of the two species seem compatible in all their
main features. In particular, all the EHV peaks identified
in CO(2--1) can be seen in SiO(2--1), especially when 
using maps with noise decreased by clipping, and
the velocity pattern of the PV diagram is identical
for both species.
This similarity between the CO(2--1) and SiO(2--1) data
is best understood if the two emissions trace the
same gas component despite their potentially
different excitation and chemical properties.

To further compare the CO(2--1) and SiO(2--1) emissions,
we fit the distribution of EHV SiO(2--1) using the same 2D
gaussian procedure used for CO(2--1). To ensure that the
results for the two species are comparable, we force the 
SiO fit to use 2D gaussians with the
same position and dimensions as those derived from CO(2--1),
so the only free parameter in the SiO fit is the peak intensity.
The results from this procedure, presented in Table 1, indicate 
that the ratio between the CO(2--1) and SiO(2--1) integrated 
intensities is close to 1 over the whole blue lobe, and that it
ranges from 1 to 4 over the red lobe, having a trend 
to increase (i.e., SiO to weaken) with distance from 
the IRAS source. Without additional information from other 
transitions, assumptions on the molecular excitation are needed 
to convert the observed intensity ratio into a ratio of column densities. 
The detection of SiO(2--1), which has a critical density of about
$4\; 10^5$ cm$^{-3}$, suggests that the EHV gas is relatively dense,
in agreement with previous analysis of similar EHV peaks in the
L1448-mm outflow by \citet{bac91a} and \citet{nis07}.  
As these authors find that the density and temperature of the gas in the 
EHV peaks typically 
ranges between $10^5$-$10^6$~cm$^{-3}$ and 20-500~K
(see also \citealt{hat99}), 
we have run a series of LVG radiative transfer models
covering that range and determining the CO(2--1)/SiO(2--1)
intensity ratio assuming optically thin emission. From this
grid of models, we find that the conversion factor between
the CO(2--1)/SiO(2--1) intensity ratio and the CO/SiO 
column density ratio is on average about 400, with
an approximate factor of 2 variation in the range of
expected densities and temperatures. We thus estimate that
the CO/SiO abundance ratio in the EHV gas is close to 400,
at least in the inner half of the jet. There is a possible increase
of this ratio at large distances in the red lobe, but we cannot
rule out that it is an effect of a density decrease in the
outer EHV, as expected from the expansion of the gas, and as
seen from a multi-line SiO analysis of the L1448-mm outflow 
by \citet{nis07}.
Observations of additional SiO transitions in
I04166 are clearly needed to clarify this issue.

SiO emission in outflows has usually been interpreted as resulting
from the release of silicon atoms in the dust grains of the ambient 
cloud due to their shocking by the outflow primary wind \citep{cas97,sch97}.
As discussed in the previous section, however,
the kinematics of the EHV gas suggests that this
outflow component is not shocked ambient material, but
gas emitted in the form of a jet from
the innermost vicinity of the central source, so
the standard explanation for SiO production seems not to
apply to this component.
An alternative SiO formation mechanism that is more 
likely to apply to the EHV gas is the one presented by
\citet{gla91}, who have carried out simplified (1D)
models of the chemistry of a primary wind from a low-mass protostar. 
These authors have shown that molecules can form efficiently via gas-phase 
reactions in an initially
atomic protostellar wind, provided that the density and
temperature of the wind stay within a range of appropriate values.
CO, in particular, forms over a wide range of density conditions,
and it tends to reach an equilibrium abundance of $4\; 10^{-4}$,
while SiO is more sensitive to density and to the possible presence
of a photodissociating far-UV radiation field. Although the results from
the \citet{gla91} model are not not fully comparable to our observations
because the model assumes a luminosity of the central source 
that is two orders of magnitude higher than the 0.4~L$_\odot$ estimated for
I04166 (and the model wind speed is probably a factor of two 
too high), they do provide an order of magnitude estimate of the
expected chemistry in the EHV component. As mentioned
previously, we estimate that the mass loss
rate in the EHV component of I04166 is of
the order of $5\; 10^{-7}$~M$_\odot$ yr$^{-1}$,
which is in the low range of values considered by \citet{gla91}.  
For this mass loss rate, the models predict a substantial production of
CO, and an abundance that depending on the details of the model
can be as high as $10^{-5}$. The expected CO/SiO abundance
ratio for this mass loss rate, however, is orders of magnitude lower
than observed, although SiO formation is so sensitive to
density that model predictions for a mass loss rate only ten times higher
can easily match our observed CO/SiO ratio.
Indeed, the gas density inside
the jet-like EHV component of I04166 is likely to be higher
than assumed by Glassgold et al. for their 
spherical models (even if these authors use a simple modification
of the density law to simulate the effect of collimation), as the 
divergence of the spherical wind ends up dominating the
density drop. In addition to this collimation effect,
a further density enhancement occurs in the IWSs, and under 
appropriate conditions, this enhancement can lead to
a higher rate of molecule formation \citep{rag05}.
Thus, gas-phase 
production in the jet seems a viable mechanism for the 
formation of CO and SiO in the EHV gas. 
A further exploration
of the chemistry of the EHV component and its comparison with the
shock-dominated chemistry of the slower outflow material will be
presented elsewhere (Santiago-Garc\'{\i}a et al. 2008, in preparation).

\section{Implications for outflow models}

Our observations of I04166 show that jet-like and
shell-like distributions of CO emission can be present 
simultaneously in a young bipolar outflow. 
Although not often emphasized, such a
coexistence of jet and shell elements in the outflow material
is not unique to I04166, and can be inferred in other
outflows from Class 0 sources with a varying degree of detail depending on
the angular resolution and signal-to-noise ratio of the
observations. The outflow from L1448-mm,
for example, has a jet-like EHV component first seen with single
dish observations by \citet{bac90}. Further higher angular resolution
data of this outflow from the PdBI \citep{bac95} have revealed that the
low velocity outflow gas moves along two opposed shells, although
these observations did not provide enough sensitivity to study in
detail the EHV component and its relation to the shells. Recent
observations of the L1448-mm outflow with the SMA by \citet{jor07} confirm
the presence of shells surrounding symmetrically the EHV component.
Additional outflows where highly collimated jets of EHV gas coexist with 
quasi-conical shells of low velocity material include the one powered
by IRAS 03282+3035, whose EHV jet was discovered by \citet{bac91b}, 
and whose low velocity shells have been recently mapped
with the OVRO interferometer
by \citet{arc06}. The HH211 outflow, observed with a number of
interferometers \citep{gue99,pal06,hir06,lee07}, also presents a high velocity 
jet and a partially-surrounding low velocity shell.
These and other observations suggest
that EHV jets are often or always surrounded by lower velocity shells, although a
systematic study of a larger sample of Class 0 outflows 
is still needed to confirm this conclusion.
If the hypothesis is correct, the I04166 outflow should be considered 
as a particularly clean example of its class, probably due to
its relatively close proximity and favorable orientation on the sky.

As discussed in the introduction, the presence of both 
jet-like and shell-like features in an outflow poses a problem
to models that assume a simple geometry for the outflow driving agent.
Neither models with only a jet-like component \citep[e.g.,][]{mas93}
nor wide-angle wind models \citep[e.g.,][]{shu91} seem capable of
explaining simultaneously the jet and shell features seen towards I04166
and similar objects.
In order to reproduce these features, a model needs to 
include simultaneously a central jet and a surrounding wide-angle wind,
both emerging from the very vicinity of the central embedded object.
In recent years, a number of models with these characteristics
have been presented in the literature, suggesting that 
two-component winds
constitute a natural geometry for a protostellar outflow.
\citet{ban06} have carried out
MHD simulations of a rotating core undergoing gravitational collapse
and found an outflow consisting of an inner jet powered by
magnetocentrifugal forces surrounded by a broad outflow driven by toroidal
magnetic pressure. Unfortunately, these simulations only extend to
600~AU, which is not much larger than the resolution element
of our observations. 
From a different MHD collapse simulation, Machida et al. (2008)
also find a two-component outflow, this time consisting of
a slow, wide-angle wind 
driven from the adiabatic (first) core which surrounds
a faster, highly collimated
jet driven from the protostar (or second core). 

Although this geometry
matches rather nicely the observations presented here, 
the outflow speeds predicted by these simulations
($\approx 5$ km s$^{-1}$ for the wide angle wind and 
$\approx 30$ km s$^{-1}$ for the jet) seem significantly 
lower than the values observed towards I04166 and similar outflows 
(the L1448-mm jet has an
radial velocity of about 50~km s$^{-1}$, as measured by
\citealt{bac90}). 

An alternative model of an outflow
with both jet-like and wide angle components is the
``unified model'' presented by \citet{sha06}.
These authors have carried out a numerical (Zeus2D) simulation
of the interaction between a protostellar wind
(based on the X-wind theory of \citealt{shu94})
and a density distribution expected for a magnetic
dense core \citep{li96},
and they have predicted the appearance of the
resulting outflow-core system as a function of time.
From these simulations, \citet{sha06} find that the gas
distribution in each outflow lobe is dominated 
by a close-to-conical shell of low velocity gas
and a highly collimated central jet of fast material,
two features that bear remarkable similarities with those
observed towards I04166.
This resemblance between the predictions from the unified model and 
the I04166 observations is not only morphological,
but kinematical, as both the jet and the shells have
velocities close to those observed in Class 0 outflows. In addition,
the shells in the unified model contain accelerated ambient gas
with a strong longitudinal component, and the
highly collimated jet represents 
the protostellar wind traveling at close to constant speed.
These two characteristics agree with the properties derived in the
previous sections for the shells and the jet of I04166, 
suggesting that the unified model captures at least some 
of the basic physics underlying the youngest bipolar outflows.
The analysis of I04166 presented here, however, indicates that 
time variability in the fastest component and the resulting generation 
of IWSs along the jet are dominant effects in the observed molecular
emission, so their inclusion in the simulations is still 
required to produce a realistic model of a bipolar outflow.

\section{Summary}

We have presented results from CO(2--1) and SiO(2--1) interferometer 
observations of the outflow powered by I04166, one of the
youngest protostars in the Taurus molecular cloud. From the analysis 
of the geometry and kinematics of the emission, together with a comparison with
existing models of outflow physics and chemistry, we have reached the
following conclusions:

1. At a resolution of about 1.5 arcsec, the outflow seems 
powered by a single YSO whose disk (plus unresolved inner envelope) 
has a mass of 0.02~M$_\odot$.

2. The bipolar outflow is highly symmetric with respect to the 
position of I04166. Each outflow lobe 
consists of two separate and well-defined components.
At radial velocities lower than 10~km s$^{-1}$, the gas 
lies along two opposed
limb-brightened conical shells that have the YSO at their vertex and
that have a full opening angle of 32 degrees.
At radial velocities higher than 30 km s$^{-1}$, the gas 
forms a pair of highly collimated jets that emerge from the vicinity of the
YSO and travel along the shell axis.

3. The geometry and kinematics of the low-velocity outflow shells 
are consistent with the gas being ambient cloud material
that has been accelerated by a wide-angle wind. In agreement with this
interpretation, we find that the northern blue shell 
coincides with the walls of an evacuated cavity seen as a reflection nebula in 
Spitzer NIR images. 

4. The highly collimated jet shows
no evidence for precession and consists of a symmetric collection of
at least 7 pairs of intensity peaks. The peaks broaden
with distance to I04166, and several of the outermost ones present
shapes reminiscent of bow shocks. The full opening
angle of the peaks is about 10 degrees, which is insufficient to
explain the low velocity shells as a result of the sideways
acceleration of the ambient gas by bow shocks in the jet.
The mass loss rate estimated (with a large uncertainty)
from this jet component is $5 \times 10^{-7}$~M$_\odot$~yr$^{-1}$.

5. The velocity field of the collimated gas presents a 
sawtooth pattern that combines a close-to-constant mean velocity
with internal gradients inside the emission peaks. In each
emission peak, the gas closer to the YSO (tail) moves faster than the
gas further away from it (head). The transition between 
these two speeds is almost linear, and it has a slope that 
flattens with distance to I04166.
Such a velocity pattern is inconsistent with the emission peaks
being solid bullets, collections of shrapnel, or shocked
ambient gas. It is
in good agreement with the predictions for internal working 
surfaces resulting from time variability in the outflow ejection
speed. Variability in the central accretion may be responsible for
these velocity changes in the outflow ejection.
The time scale of this variability is of the order of 100 years.

6. The geometry and kinematics of the highly-collimated, fast gas  
suggests that this component consists of material emitted from
the protostar or from its immediate vicinity, and not of swept
up ambient gas. The relative abundance of CO and SiO that we derive 
for this component is in reasonable agreement with the chemical
model of a protostellar wind presented by \citet{gla91}.

7. The combination in a single outflow of low-velocity
shells accelerated by a wide-angle wind 
and a fast, jet-like component moving along the shell axis
illustrates the need for an outflow mechanism that
produces these two different features simultaneously. 
The recent ``unified" model of \citet{sha06}
seems in good agreement with these characteristics, 
although
the presence of internal working surfaces in the
I04166 outflow indicates that time dependence 
in the ejection velocity is an additional element 
needed to model realistically outflow observations.


%

\begin{acknowledgements}
We thank Arancha Castro-Carrizo, Jan Martin Winters, and Aris Karastergiou for
help preparing and calibrating the PdBI observations, Fr\'ed\'eric Gueth
for advise on combining single-dish and interferometer
observations, Bringfried Stecklum for information on H$_2$
emission around I04166, and Antonio Crapsi for carrying
out radiative transfer models of the dust continuum emission 
from I04166.
We also thank the referee, John Bally, for a number of useful comments 
that helped clarify the presentation.
JS-G thanks Sienny Shang for her hospitality during a visit to the
ASIAA and for valuable information on the unified outflow model.
JS-G, MT, and RB acknowledge partial support from project 
AYA 2003-07584.
This research has made use of NASA's 
Astrophysics Data System Bibliographic Services and the SIMBAD 
database, operated at CDS, Strasbourg, France. 
This work is based in part on observations made with the Spitzer 
Space Telescope, which is operated by the Jet Propulsion Laboratory, 
California Institute of Technology under a contract with NASA.
\end{acknowledgements}

\end{document}